# The Reception of the Copernican Revolution Among Provençal Humanists of the Sixteenth and Seventeenth Centuries[*]

Jean-Pierre Luminet
*Laboratoire d'Astrophysique de Marseille (LAM) CNRS-UMR 7326
& Centre de Physique Théorique de Marseille (CPT) CNRS-UMR 7332
& Observatoire de Paris (LUTH) CNRS-UMR 8102
France*
E-mail: `jean-pierre.luminet@lam.fr`

## Abstract

We discuss the reception of Copernican astronomy by the Provençal humanists of the XVI[th]-XVII[th] centuries, beginning with Michel de Montaigne who was the first to recognize the potential scientific and philosophical revolution represented by heliocentrism. Then we describe how, after Kepler's *Astronomia Nova* of 1609 and the first telescopic observations by Galileo, it was in the south of France that the New Astronomy found its main promotors with humanists and « amateurs écairés », Nicolas-Claude Fabri de Peiresc and Pierre Gassendi. The professional astronomer Jean-Dominique Cassini, also from Provence, would later elevate the field to new heights in Paris.

## Introduction

> *In the first book I set forth the entire distribution of the spheres together with the motions which I attribute to the earth, so that this book contains, as it were, the general structure of the universe.*
> —Nicolaus Copernicus, Preface to Pope Paul III, *On the Revolution of the Heavenly Spheres*, 1543.[1]

Written over the course of many years by the Polish Catholic canon Nicolaus Copernicus (1473–1543) and published following his death, *De revolutionibus orbium cœlestium* (On the Revolutions of the Heavenly Spheres) is regarded by historians as the origin of the modern vision of the universe.[2] The radical new ideas presented by Copernicus in *De revolutionibus* first appeared some three decades earlier in a brief manuscript, the *Commentariolus*, but with a circulation limited to a small group of intellectuals.[3]

---

[*] Extended version of the article "The Provençal Humanists and Copernicus" published in *Inference*, vol.2 issue 4 (2017), on line at http://inference-review.com/. Translated from French by the editors.

In the *Commentariolus* and *De revolutionibus*, Copernicus sought to replace the geocentric thesis enshrined by Aristotle some two thousand years earlier. Reinforced by the Alexandrian astronomer Claudius Ptolemy (c.100–c.170 CE) in his *Almagest*, the culmination of ancient Greek observational and mathematical science, the geocentric thesis had dominated western and Arab astronomy throughout the medieval period.

Aware of imperfections in Ptolemy's system, and eager to discern a new order in the cosmos, Copernicus proposed an astronomical model in which the sun was at the geometric center of the universe, while the earth orbited it over the course of a year, and rotated on its own axis over the course of a day. A very similar model had been suggested by Aristarchus of Samos in antiquity. Heliocentrism demoted the earth to the status of a planet, another wandering star, just like Mercury, Venus, Mars, Jupiter, and Saturn. Our planet no longer occupied a cosmologically privileged position.

In the second half of the sixteenth century, the first in France to mention Copernicus was Omer Talon (1510–1562), a disciple of Petrus Ramus, in his *Academicae questiones* (1550). His position was quite favorable toward Copernicus; followers of Ramus were, generally speaking, hostile to Aristotle.[4] But few of his contemporaries took Copernicus seriously, and most viewed heliocentrism negatively. One can even detect some mockery of Copernicanism in the poets of *La Pléiade*, such as Jean Bodin.[5] Or, more prominently, in the following example from Guillaume de Salluste du Bartas:

*Even so some brain-sicks live there now-a-days*
*That lose themselves still in contrary ways,*
*Preposterous wits that cannot row at ease*
*On the smooth channel of our common seas.*
*And such are those, in my conceit at least,*
*Those clerks that think—think how absurd a jest!—*
*That neither heavens nor stars do turn at all,*
*Nor dance about this great, round earthly ball,*
*But th'earth itself, this massy globe of ours,*
*Turns round about once every twice twelve hours.*[6]

The apparent absurdity in claiming that the earth was not, in fact, immobile meant that the Copernican doctrine was slow to spread and slower still to gain acceptance. Although now widely termed a scientific revolution, this description did not appear until the twentieth century, when it was coined by Thomas Kuhn.[7]

## Skepticism and Astronomy

Michel de Montaigne was an exception to the general view that Copernicanism was absurd. In his *Essais*, he not only supported the heliocentric theory, but also perceived that the work of Copernicus was indeed a scientific revolution in the making. To understand the reasoning behind his embrace of heliocentrism, one must recall Montaigne's fundamentally skeptical position with respect to the philosophy of knowledge.

Montaigne received a humanist education from a very young age. He forged a career as a magistrate, served as mayor of Bordeaux, and retired in 1571 at the age of thirty-seven to devote the remainder of his life to writing and revising his *Essais*, which are comparable to thought experiments. Montaigne's views concerning the Copernican system appeared in a chapter entitled "An Apology for Raymond Sebond," in which he described with satisfaction the decline of geocentrism:[8]

*For three thousand years the skies and the stars were all in motion; everyone believed it; then Cleanthes of Samos, or according to Theophrastus, Nicetas of Syracuse decided to maintain that it was the Earth which did the moving, revolving on its axis through the oblique circle of the Zodiac; and in our own time Copernicus has given such a good basis to this doctrine that he can legitimately draw all the right astronomical inferences from it.*[9]

Montaigne adopted a broadly skeptical position against certitude in scientific matters.[10] The heliocentric model, he noted, should not be accepted simply because it was true, but rather because it dethroned man from his central place in the universe. He preferred the new theory for ethical reasons, noting that Epicurean and Stoic philosophers had favored the heliocentric model long before Copernicus, and that they had done so with an eye to questioning the importance that man had spontaneously attributed to himself.[11]

At the end of the sixteenth century, Copernicanism faced cultural and religious resistance from many thinkers. Montaigne was not among them. A fervent admirer of Lucretius's *De rerum natura* (On the Nature of Things), Montaigne had contemplated the possibility of a plurality of worlds, a view following naturally from atomistic philosophy. He considered it more likely than the view held by Aristotle and, in particular, Thomas Aquinas, that the world was unique. The plurality of worlds would, in effect, make the earth disappear into the immensity of the universe, and reinforced for Montaigne the de-centering of the earth.

According to Marc Foglia, Montaigne endorsed the Copernican hypothesis

*as a philosophical adversary of an institutionalized Aristotelianism that had become "religion and law," and as a defender of the free exercise of thought. In contrast to a political order, science is not a set of truths that one must defend, but only a dominant tradition that one should be able to question critically.*[12]

Montaigne may have read the anonymous preface to *De revolutionibus*, which attempted to neutralize the revolutionary implications of the book by depriving the heliocentric theory of any physical significance:

*For these hypotheses need not be true nor even probable. On the contrary, if they provide a calculus consistent with the observations, that alone is enough. ... Therefore alongside the ancient hypotheses, which are no more probable, let us permit these new hypotheses also to become known.*[13]

The preface is now thought to have been written by Andreas Osiander, a Protestant theologian sympathetic to Martin Luther and Philip Melanchthon, both of whom had previously accused Copernicus of contradicting the Bible.

Book I of *De revolutionibus* begins with an author's preface, written in the form of a dedication to Pope Paul III, in which Copernicus provides physical arguments in support of his model. These had been affirmed beforehand in the *Narratio prima de libris revolutionum Copernici* (The First Report on the Book of the Revolutions of Copernicus), a trial balloon of sorts for heliocentrism, published in 1540 by the sole pupil of Copernicus, Georg Joachim Rheticus (1514–1574).

One can recognize in Osiander's preface the interpretation known as instrumentalism, according to which astronomy only uses mathematical fictions in order to explain the planetary trajectories; its sole aim is to save the phenomena. This Platonically-inspired astronomy—as opposed to Aristotelian physics, which sought the causes of things—did not offer opinions on the nature of celestial phenomena.

Montaigne may have adopted this view in his *Essais*, but it was not to minimize the epistemological importance of Copernicus. On the contrary, Montaigne sought corroboration for his argument that human knowledge is illusory. The vision of the world suggested by the heliocentric theory certainly seemed better regulated and more harmonious than the geocentric theory of Ptolemy, with its complex system of epicycles. That did not establish its truth. In science, as in every discipline, Montaigne believed, one must reserve judgment, and it was thus natural to exercise caution regarding the Copernican system. One had to consider it simply as a recent, and therefore interesting, step toward a systematic description of the universe. It was thus necessarily provisional, inevitably to be replaced, sooner or later, by a better system:

*What should we take from this, if not that we should not take either one? And who knows whether a third opinion, a thousand years from now, would not reverse the two preceding ones?*[14]

As it turned out, a thousand-year wait was unnecessary. New systems were proposed shortly thereafter by Tycho Brahe in 1583 and by Johannes Kepler in 1596.

## The Galilean Revolution

Over the six decades that followed the publication of *De revolutionibus*, only a handful of astronomers throughout Europe understood the importance of the Copernican theory and sought to defend it, let alone adopt and improve it. Among them were William Gilbert and

Thomas Digges in England, Galileo Galilei in staunchly Catholic Italy, and Rheticus, Michael Maestlin, Christophe Rothmann, and Kepler in Lutheran countries. They too had to defend themselves from virulent critiques of the "absurd" doctrine of the double motion of the earth. Adopting the argument provided by Osiander, most scholars of the time considered the Copernican system to be an ingenious mathematical fiction that facilitated and improved the calculation of celestial ephemerides. This could be seen in the Prutenic Tables, calculated by Eramus Reinhold using the heliocentric model and published in 1551. These proved slightly superior to the Alphonsine Tables of 1483, which were calculated using Ptolemy's geocentric model. This general sense of mistrust towards Copernicanism was shared by Tycho Brahe (1546–1601), the most celebrated astronomer of his time and an experimentalist renowned for the quality of his observations.[15] While Tycho admired Copernicus' work, he did not support geokineticism and in 1583 proposed a geo-heliocentric model in which the earth was immobile, orbited by the sun, moon, and fixed stars, while the five planets were in orbit around the sun. Tycho was thus able to remain faithful to both the principles of Aristotelian physics and a theological interpretation of the biblical account. This clever and comforting compromise quickly attracted the support of most astronomers, philosophers, and theologians of the time, whether Catholic or Reformed.

In 1610, Galileo, who had previously not dared teach Copernican astronomy, published *Siderius nuncius* (Starry Messenger), in which he revealed the results of his observations made using a telescope. These findings contradicted the dogmas of Aristotelian physics: the moon had an uneven surface, just like earth; the sun, covered in spots, was imperfect; Venus had phases; Jupiter was at the center of a system with four moons; and there were many more fixed stars than could be seen by the naked eye. Galileo's observations had implications that extended far beyond astronomy. The disruptive doctrine of Copernicus was elevated to center stage, especially when Galileo, in his public correspondence, began defending heliocentrism from theological attacks.

At first, the ecclesiastical authorities reacted with a judicial procedure against Galileo. In February 1616, the Copernican theory was adjudged false, because it was contrary to scripture. Amendments and corrections were thus required to render *De revolutionibus* inoffensive in the eyes of the church. Robert Bellarmine, a powerful Cardinal, insisted that Galileo abandon heliocentrism, but did not forbid him to study the Copernican model as a scientific hypothesis, albeit one that had no basis in truth; a determination that could only be made by the theologians of the Holy Office.

In the years that followed, Galileo appeared relatively docile. He avoided controversy until 1632 when he published *Dialogo sopra i due massimi sistemi del mondo* (Dialogue Concerning the Two Chief World Systems), in which he mocked the Aristotelian conception of the universe, the geocentric theory that followed from it, and those who believed in it.

In adopting the heliocentric system, Galileo did not offer a hypothesis, he affirmed a reality. Ruling that he had betrayed the terms of the agreement from earlier judicial proceedings, on June 22, 1633, the Congregation of the Holy Office required him to recant

*of having believed and held the doctrine—which is false and contrary to the sacred and divine Scriptures—that the Sun is the center of the world and does not move from east to west and that the Earth moves and is not the center of the world.*[16]

In a decree dated August 23, 1634, the *Dialogo* was placed on the *Index Librorum Prohibitorum* (List of Prohibited Books)—as had been the case with the *Epitome Astronomiae Copernicanae* (Abbreviation of the Copernican Astronomy) by Kepler, which was condemned in 1619.

Under siege in it's birthplace, the prospects for Copernicanism appeared uncertain. Galileo was now working under the shadow of the Inquisition. In Prague, the Holy Roman Empire having been ravaged by the Thirty Years War, Kepler lost his post as Imperial Mathematicus. In France, René Descartes (1596–1650), made cautious by the persecution of Galileo, decided not to publish *Traité du monde et de la lumière* (Treatise of the World and of Light) in his lifetime. This may have been a prudent decision. Descartes defended the heliocentric theory, going so far as to propose an infinite space in which each star is the center of a "vortex of ether" similar to our solar system.

Amidst these difficulties, it was in the South of France that Copernican-Galilean astronomy would find new adherents during the seventeenth century. Namely, the two great Provençal humanists, Nicolas-Claude Fabri de Peiresc (1580–1637) and Pierre Gassendi (1592–1655). The astronomer Jean-Dominique Cassini (1625–1712), also from Provence, would later elevate the field to new heights in Paris.

**Peiresc, Prince of the Curious**

Nicolas-Claude Fabri de Peiresc was born December 1, 1580 in Belgentier, a small town between Aix and Toulon. The outline of his life is known from a biography written by his close friend Gassendi. As a young man, Peiresc studied with the Jesuits in Avignon, and then Tournon. At the age of sixteen, he began the study of astronomy, which he found fascinating. At the time, astronomy involved making an inventory of the stars and tracking their movements by measuring the angles between them using a cross-staff or an astrolabe. Peiresc returned to Aix-en-Provence, before moving to Padua to study law, while also engaged in many other studies. Upon his arrival in Padua, he quickly befriended the Italian humanist Gian Vincenzo Pinelli (1535–1601), who became his teacher and role model. It was from Pinelli, whose library was the largest of the sixteenth century, that Peiresc acquired his appetite for books and his interest in various cabinets of curiosities.[17] It was at Pinelli's home, too, that he met Galileo for the first time; Pinelli had opened his library to him.

After a little more than three years spent in Italy and following the death of Pinelli, which affected him deeply, Peiresc returned to France to continue his law studies. He completed his doctoral thesis in Montpellier and then, after various trips to Paris, London and Flanders, he was appointed an advisor to the Parliament of Provence. Alongside his other activities, astronomy remained one of his major concerns throughout this period. In truth, he never really left the discipline and, at certain points in his life, would devote all of his energy to it.

In the fall of 1604, Peiresc observed the meeting of the three superior planets—Mars, Jupiter and Saturn—an event known as the Great Conjunction that occurs only once every eight hundred years. Simultaneously, a star the magnitude of Jupiter appeared and was observed for more than a year in one of the legs of the constellation Ophiuchus. Peiresc lacked a celestial globe that could provide certainty regarding the number of fixed stars, and he initially believed this particular star must be among those listed by the Ancients. From letters he received a few months later, he learned that it was, in fact, a new star, one that had also been observed by Galileo, and whose advent was an additional blow to the Aristotelian conception of an immutable sphere of the fixed stars.[18]

In 1608, the field of observational astronomy was shaken by a new discovery. At the beginning of the century, Dutch opticians realized that glass lenses, which had been used since the thirteenth century to correct vision, could be configured to magnify distant objects. 0The telescope was invented.

Upon hearing of these developments, Galileo immediately began constructing an instrument with a magnification power of thirty times, the device that now bears his name. In contrast to his contemporaries, for whom the observation of distant territorial objects using the new device had become an increasingly popular pastime, Galileo pointed his telescope toward the heavens. In November 1609, he observed and described the lunar terrain, sunspots, and the phases of Venus. On January 7, 1610, he discovered four new planets circling Jupiter, which he named the Medicis.[19] This was the beginning of modern astronomy.

Peiresc was informed of Galileo's new discoveries in a letter dated May 3, 1610. Quickly grasping the importance of the new technology, he began work on an instrument of his own. In November 1610, Peiresc, equipped with his new device, started making observations from his terrace, which he had prepared for the occasion. He was joined by a group of amateur astronomers, including Joseph Gaultier de la Valette (1564–1647), Vicar General of Aix. On November 24, 1610, Peiresc and Gaultier were the first in France to observe the four satellites of Jupiter, and on November 26 they discovered the Orion Nebula, which Peiresc described as follows:

> *In Orione media ... Ex duabus stellis composita nubecula quamdam illuminata prima fronte referabat coelo not oio sereno. (At the center of Orion ... a cloudiness composed between two stars and in some sense seen from the front and illuminated from behind, the sky not being perfectly clear.)*

Peiresc dedicated himself to the observation of the Galilean satellites. He named them Cosmus Minor (Callisto), Cosmus Major (Ganymede), Maria (Europa), and Catharina (Io). His group of amateur astronomers followed their movements, measuring the lengths of their orbits around the planet and their disappearances behind it. They soon realized that these punctual and frequent eclipses could be helpful in determining longitudes, and to this end they constructed tables that predicted the positions of the satellites at specific times. Upon learning that Galileo was working on the same problem, Peiresc elected not to publish his results and generously gave up his project out of respect for the elder scholar, whom he greatly admired. As it turned out, the predictions made by Peiresc's group in Aix would prove more accurate than those of Galileo.

The orbital periods of the four Galilean satellites of Jupiter were estimated by Peiresc with great precision:[20]

| Satellite | Peiresc's Value (Days) | Current Value (Days) |
|---|---|---|
| Io (Catharina) | 1.7 | 1.769 |
| Europa (Maria) | 3.5 | 3.551181 |
| Ganymede (Cosmus Major) | 7.14 | 7.15455296 |
| Callisto (Cosmus Minor) | 16.7 | 16.6890184 |

Peiresc's greatest innovation was a graphical scheme in which he linked the successive positions of Jupiter, two of its satellites, and some bright stars by means of a sinusoidal curve. His diagrams reveal the first use of graph paper in modern terms, but more to the point, they reveal an astronomer prepared to track the motions of a heavenly body in four dimensions rather than three. This allowed Peiresc to determine the positions of the satellites for dates when poor weather had prevented observations, and to predict the positions they would occupy in the future. The curves described by Peiresc are still used today to prepare observations of Jupiter's satellites, although they did not appear in the *Annuaire du Bureau des longitudes* until a century later.

Whatever Peiresc may have thought about Jupiter and its satellites, the Inquisition understood the implications of his work only too well. With moons in regular orbit around Jupiter, neither going astray nor disturbing the path of the planet, there was nothing, in theory, to prevent the motion of the earth around the sun.

The Inquisition had by now condemned Galileo; he found himself under house arrest at his villa in Arcetri, near Florence. Peiresc was torn. He was not a cleric, but he was almost a man of the church; the pope and numerous cardinals knew him and respected his immense knowledge. Peiresc's options were limited and, at best, he could only attempt to relieve Galileo's misfortunes. In his correspondence with Francesco Barberini, a cardinal and nephew of the pope, Peiresc assumed the role of Galileo's advocate (in a letter dated December 5, 1634): "*My hope is that you will deign yourself to do something for the consolation of a good old septuagenarian, who is in ill health and whose memory will be difficult to erase in the future.*"[21] His pleas had no effect. Peiresc continued his efforts without success.

At the same time, he returned to the problem of determining longitudes. The best way to measure a difference in longitude was to observe a celestial event in two distant places; the time difference between observations is a measure of their longitude. No clocks were at the time both reliable and portable. Observing the satellites of Jupiter was not a task that could be easily carried out by sailors, for example, but lunar eclipses lent themselves more readily to this type of measurement.

With this in mind, Peiresc sought to arrange for the coordinated observation of a lunar eclipse by observers distributed along the Mediterranean. His correspondents were numerous; most were friars, since he had obtained the agreement of the leaders of the Jesuits and the Dominicans. After years of organizing, on August 28, 1635, the participants—in Aix, Aleppo, Cairo, Digne, Naples, Padua, Paris, and Rome—were ready to determine the local time at which the moon moved into the shadow of the earth. The instructions and the instruments did not arrive in Tunis in time for the eclipse. Mt. Sainte-Victoire, which towers over Aix-en-Provence, was climbed by a group of observers who missed the event because they were asleep!  The results of the project were not insignificant; determining the longitude of Aleppo also yielded a new figure for the length of the Mediterranean, one that was almost a thousand kilometers shorter than previous estimates.

Yet Peiresc was not entirely satisfied. It was not when the moon entered or exited a shadow cone that one should make an observation, he noted. One had to fix on more precise points. For this purpose, a map of the moon's surface would provide recognizable terrain. With the backing of Gassendi—whom he called the prince of the curious—Peiresc asked the engraver Claude Mellan to prepare the first detailed maps of the moon. These were based on telescope observations made from the personal observatory that Peiresc had built on the roof of his house. Two spectacular charts were engraved in 1636, but Peiresc's death the following year prevented the task from being completed. The crater Peirescius, close to the Mare Australe in the moon's southeastern hemisphere, is named in honor of Peiresc.

Peiresc collected his own astronomical observations and those of his contemporaries in a series of manuscripts. He included numerous graphs and calculations, tables of ephemerides, and pages with ideas that had never appeared in published form—making Peiresc one of the great forgotten men in the history of astronomy.[22] He also republished a selection of letters he had received on various scientific subjects. A tireless correspondent, he composed thousands of letters during his lifetime, most of which are now preserved at the Inguimbertine library in Carpentras and the Méjanes library in Aix-en-Provence. Peiresc's most frequent correspondent was his dear friend Gassendi, but as a good humanist, interested in all the noble disciplines, he also corresponded with, among many others, the poet François Malherbe, the painter Rubens, Galileo, and the philosopher Tommaso Campanella, whom he also attempted to defend from the attacks of the Inquisition.

Peiresc died in 1637. Galileo, a prisoner of the Holy Office, went blind. Even though their work had been challenged, scientists had clarified how the planets orbited the sun, all within a celestial vault for which observational instruments were providing a better and better inventory. The results as a whole would lead to the implacable mechanics that govern motion: the law of universal gravitation, discovered by Isaac Newton, who was born almost exactly one year after the death of Galileo in 1642.

The scientific world's tributes to Peiresc have, rightly, not been limited to naming a lunar crater in his honor. The asteroid 19226, discovered in October 1993, was also named after him, a rather belated recognition of his contributions.

## Gassendi, from Astronomy to Atomism

Pierre Gassendi was born January 22, 1592, near Digne, in the Alps of Haute-Provence. After beginning his studies in Digne, he took a course in philosophy at the University of Aix. In 1614, after obtaining a doctorate in theology at Avignon, Gassendi was named Professor of Rhetoric and canon at Digne, and later Professor of Philosophy at Aix, where he seems to have been chased by the Jesuits. He died on October 24, 1655, in Paris, after having been named, ten years earlier, Professor of Mathematics at the Royal College (later the *Collège de France*). Gassendi was the very model of a polymath humanist; he was an astronomer, mathematician, philosopher, theologian, and writer. In addition to his biography of Peiresc, Gassendi also left us invaluable biographies of the astronomers Copernicus, Kepler, Regiomontanus, and Tycho.[23] But it was in astronomy and philosophy, in particular, that his legacy would prove most enduring.

Born to a family of poor farmers, in his youth Gassendi discovered his passion for the wonders of the heavens while watching over the flocks at night. He made regular observations throughout his life, using both telescopes and sighted instruments. During the first half-century following the invention of the telescope, the two methods were used in parallel. With telescopes one sought to make discoveries, while with traditional instruments, such as the quadrant or Jacob's staff, one made measurements, something that could not yet be done with telescopes (the micrometer, developed by William Gascoigne in 1639, was not widely known at the time).

Sunspots were one of the great novelties revealed by the telescope. Were the spots on the surface of the sun, or were they small satellites in orbit around the sun? Were they clouds, or even imperfections in the telescope itself? Gassendi began a long series of observations in 1620, increasing in frequency around 1626. This was the year that Christoph Scheiner (1575–1650) published his first works on the question, proposing that the spots were, in fact, satellites. Gassendi, for his part, was in agreement with Galileo, who suggested that the spots were marks on the surface of the sun itself, and thus proof for the rotation of our star.

From his observation of sunspots, Gassendi determined a speed of rotation for the sun, obtaining an estimate of 25 to 26 days, a fairly remarkable result given the values that varied according to latitude.

Unfortunately, most of Gassendi's solar observations are lost because they were made before he began keeping notes in his journal systematically. He later became one of the first astronomers to understand the importance that a collection of observations could have. On September 27, 1635, Gassendi wrote to Peiresc that:

*[I]n order to prevent these scribbles and notes from going astray, I have begun for some while to write everything into a whole quire of paper, which I have sewn and covered with parchment for this purpose.*

His diary, or astronomical journal, was born at the same time as his recognition of the essentially historical nature of astronomy, which motivated him to organize and preserve his own observations.

I mentioned above the project to create a lunar atlas, a joint project undertaken by Gassendi and Peiresc, beginning in 1634. From September to December 1636, one can follow the lunar observations in the diary of Gassendi. Alas, the death of Peiresc on June 24, 1637 put an end to the preparation of the atlas. Mellan remained in Paris, and Gassendi, deeply affected by the passing of Peiresc, abandoned the project. As he explains in his biography of Peiresc, their goal, aside from pure astronomical interest, was the cosmological order.[24] They hoped to demonstrate that the lunar globe was similar to the terrestrial globe, and to offer support for Galileo's intuition about the deep unity of terrestrial and celestial physics.

Peiresc and Gassendi were not the first astronomers to chart the moon. They followed in the footsteps of Galileo, as did others such as Scheiner, Guiseppe Biancani, and Thomas Harriot, who made the first known sketch of the moon on April 5, 1609. But Peiresc and Gassendi were the first to compile a complete lunar atlas, or selenography. In doing so they were following in the tradition of Galileo's disciples who were anxious to prepare a detailed inventory of the sky using the telescope, and to apply it to the practical needs of daily life.

In Gassendi's diary, one can see that he observed the planets as best he could. For Mars, he was engaged in determining its angular distance to the stars, and, for Jupiter, he pursued the program of observing the satellite undertaken by Peiresc. He was also interested in the strange shifting form of Saturn, not suspecting the existence of the planet's rings (in 1655, the year of Gassendi's death, the Dutch astronomer Christiaan Huygens (1629–1695) would be the first to describe the rings of Saturn as a disc).

Of all the planetary observations made by Gassendi, the most important was without a doubt that of Mercury passing in front of the Sun on November 7, 1631. For a planet to transit the solar disc, it must pass between the earth and the sun. Only Mercury and Venus can be observed from the earth during their transit. Similar passages took place in the months of May and November, around the 7th and the 9th of the month. For Mercury, one could expect this phenomenon to repeat every 7, 13, or 46 years. It was thus a rare observation, which explains its interest for astronomers.

The passage of Mercury before the sun on November 7, 1631 was the first transit to have been predicted and conscientiously observed. One reason for this was the difficulty of knowing exactly when the event would take place. The tables astronomers possessed at the beginning of the seventeenth century were quite unreliable. In the ephemerides that Kepler had calculated for the years 1629 to 1631—on the basis of the *Rudolphine Tables* of 1627—he had added a note, titled *Admonitio*, indicating that Mercury would transit the sun on November 7, 1631. Following Kepler's death in 1630, his son Jacob Bartsch had the note republished as an offprint.

Gassendi, like other astronomers in Europe, read Kepler's booklet and made his preparations. The pair alerted their fellow astronomers in Provence to the forthcoming event. Since Peiresc had not read Kepler's *Admonitio*, Gassendi wrote him a long letter on July 9, 1631, with the information and instructions. Gassendi was in Paris at the time and did not expect to see much from a northern latitude in the middle of November. He was counting on Peiresc's observers and clearer skies in Provence.

November arrived and the time when Mercury was to traverse the solar disk. Neither Peiresc nor Gaultier, nor any other member of the Provençal group saw anything. In fact, it was Gassendi in Paris who, alone in France, made the observation!

Gassendi took great pains with his observation. Since he could not look directly at the planet, he had the idea of projecting its image onto a sheet of paper. On November 5 he began his vigil, despite steady rain throughout the day. The following day he saw the sun only briefly through a continuing downpour. But on the 7th, the sun could be seen clearly for periods of time. Mercury was already visible on its surface, though Gassendi had difficulties in recognizing it due to the small size of the black spot. He published an account of his observations in a pamphlet entitled *Mercurius in sole visus* (Mercury Seen in the Sun).[25]

Gassendi may have been shocked by the relatively tiny size of Mercury, but his observations confirmed Galileo's predictions that the planets were much smaller than they seemed, and indeed smaller than astronomers had previously thought. Above all, Gassendi's observations reinforced the authority of Kepler's Rudolphine Tables, and in a more general sense confirmed the efficacy of the new astronomy. Equally, it obliged astronomers to reexamine the question of stellar and planetary diameters, and thus their distance with respect to the earth and the sun.

Throughout his career as an experimental astronomer, Gassendi rarely missed observing an eclipse of the sun or moon, describing his pursuit of them as "like a cat after mice." The list of his observations, made over a period of thirty years, is lengthy.

The extreme care with which Gassendi always worked marks an essential stage in the development of observational astronomy. That the second half of the seventeenth century saw the appearance of an astronomy with a level of precision far beyond that which had preceded it is, in part, a consequence of the techniques employed and improved by Gassendi, and passed on by him to the next generation.

The course that Gassendi taught at the Royal College, edited and published under the title *Institutio Astronomica* (1647), became a respected manual in France, England, Italy, and America. Gassendi was identified with the new astronomy due to his practical work, his biographies of Copernicus, Tycho, and Kepler, and his observation of the transit of Mercury. Re-edited many times, his manual appeared in public and private libraries alongside the other foundational texts of modern and revolutionary astronomy by Copernicus, Galileo, and Kepler.

In philosophical terms, Gassendi can be categorized among the atomists, and in particular among the Epicurians, whom he helped rehabilitate; a rather surprising position for an ecclesiastic. He was long opposed to the official theories of Aristotelianism and supported

Galileo in his denunciation of geocentrism. He was equally opposed to any form of obscurantism, especially astrology.

The works of Gassendi had a particular importance for Italian scientists. Monitored by an authoritarian and reactionary church, experimenters and empiricists intimidated by the condemnation of Galileo were, like Gassendi, in search of a philosophical system that could explain and order the facts provided by the fragmentary experiments of the time. Gassendi's system—namely, the ancient atomism of Epicurus purged of its atheistic tendencies—offered the Italian intelligentsia an alternative to neo-Aristotelianism without heading in the direction of the deterministic mechanism of Descartes, with whom Gassendi was engaged in a long epistolary dispute. For them, Gassendi had become the natural complement to Galileo.

Unlike Giordano Bruno in his *De innumerabilibus, immenso, et infigurabili* (Of Innumerable Things, Vastness, and the Unrepresentable), Gassendi did not go so far as to defend the idea of a plurality of worlds. This was a theory with unacceptable theological implications and the unfortunate Bruno was tried for heresy by the Inquisition and burned at the stake in 1600. Gassendi owned a copy of *De innumerabilibus* and in his own writings noted that he was in agreement with some aspects of Bruno's theories, such as the idea that the stars are, in fact, themselves suns, each with their own orbiting planetary systems. Gassendi was a follower of Digges in this respect, rejecting the notion that the universe is enclosed within a spherical shell of fixed stars, an idea that had persisted since Antiquity and was not contested by Copernicus in *De revolutionibus*. Digges proposed that the stars were instead scattered throughout the universe at varying distances.[26]

The most important influence that the writings of Gassendi exerted in England was on John Locke and Isaac Newton. The impact of the French humanists on Locke's *Essay on Human Understanding* (1690) was noted by Leibniz:

*This author is pretty much in agreement with M. Gassendi's System, which is fundamentally that of Democritus: he supports vacuum and atoms, he believes that matter could think, that there are no innate ideas, that our mind is a tabula rasa, and that we do not think all the time; and he seems inclined to agree with most of M. Gassendi's objections against M. Descartes.*[27]

## From Provence to Paris : Cassini the First

I will conclude this essay by briefly reflecting on the life and work of the astronomer Gian-Domenico Cassini (1625–1712), also of Provençal origin : he was born in Perinaldo, which at the time was part of the region of Nice in the Duchy of Savoy. Educated in the Jesuit school of Genoa, his talents brought him to the attention of a wealthy amateur from Bologna, the Marquis Cornelio Malvasia. In 1644, the Marquis engaged Cassini to work at the Observatory of Panzano, which was still under construction at the time. Numerous instruments were put at his disposal and he worked with the Jesuit fathers Giovanni Riccioli and Francesco Grimaldi, two astronomers of great repute, who completed his education.

The quality of Cassini's observations and his astronomical publications earned him the honor of being named Professor of Astronomy and Mathematics at the Jesuit University of Bologna in 1650. He was then just twenty-five years old.

In the statutes under the jurisdiction of the Roman Catholic Church, Cassini was obliged to teach Ptolemaic astronomy. Nevertheless, after observing the comet of 1652–1653, he adopted Tycho's geo-heliocentric system, already favored by the Jesuits; he would not accept the Copernican model until later.

An expert in both hydraulics and engineering, Cassini acquired such a reputation that the senate of Bologna and the Pope charged him with a number of scientific and political missions. But it was astronomy to which he devoted most of his energies. In 1665, Cassini discovered Jupiter's giant red spot and determined the rotation speed of Jupiter, Mars, and Venus. His reputation had by now spread beyond the borders of Italy and in 1668, Colbert, who sought out foreign scientists for the new *Académie des Sciences* in Paris, asked him to become a corresponding member. Cassini accepted.

Colbert then invited Cassini to come to France for a limited period of time to help him construct a new observatory. Cassini arrived in Paris in August of 1669 and soon began collaborating at the *Académie*, modifying the plans of the architect Perrault in order to better adapt the building for astronomical observations. Considered the finest astronomer of his time, he was named director of the Observatory of Paris at the request of Louis XIV, and charged with making it the most important astronomical and scientific center of the time, a goal which he attained.

Even before the Observatory was completed, Cassini had begun his observations and research, discovering two of Saturn's satellites: Iapetus in 1671 and Rhea in 1672.

Despite requests from the Pope, Cassini expressed his desire to remain in France and requested citizenship, which he obtained in 1673. He then Gallicized his first name to Jean-Dominique.

The same year, Cassini made the first precise measurement of the distance from the earth to the sun, using a measurement of the parallax of Mars deduced from the observations of Jean Richer in Cayenne. Two years later, he discovered the division of the rings of Saturn, now known as the Cassini Division, and, in 1684, two new satellites of Saturn: Tethys and Dione. In 1679, Cassini presented a cartography of the moon to the *Académie des Sciences* that surpassed Mellan's maps and would be unequaled in precision until the invention of photography. Around 1690, he was also the first to observe the differential rotation in the atmosphere of Jupiter. Between 1668 and 1693, he published a great deal of material, including his *Ephemerides Bononienses mediceorum siderum* (Ephemerides of the Satellites of Jupiter). In 1676, at the Paris Observatory, Ole Rømer (1644–1710) demonstrated for the first time that the speed of light was not infinite. Rømer estimated that, *"light seems to take about ten to eleven minutes [to cross] a distance equal to the half-diameter of the terrestrial orbit."*[28] Oddly enough, Cassini refused to recognize Ole Rømer's demonstration, despite the fact that Rømer used Cassini's own ephemerides of Jupiter's satellites in his calculations.

Cassini went blind in 1710 and died two years later in Paris at the age of 87. Succeeded by his son, Jacques Cassini, three generations of Cassinis reigned at the Observatory of Paris over 122 years. Jean-Dominique Cassini would be designated by the regal title Cassini I$^{er}$.

## The Adventurers

The radical change in our conception of the cosmos that began with Copernicus and was completed in 1687 with the publication of Isaac Newton's *Philosophiæ Naturalis Principia Mathematica* (Mathematical Principles of Natural Philosophy) is the logical extension of the episodes recounted here, of which those concerning Peiresc and, to a lesser extent, Gassendi, however much they may figure in scholarly accounts, are not well known to the public. They have remained in the shadows. Yet it is these intellectual adventurers who continued the revolution begun in the previous century and helped pave the way for the modern world. Not content with revolutionizing astronomy and science, the new philosophy derived from Copernican astronomy, powerfully developed by Kepler and Galileo and propagated in Europe by the Provençal humanists of the seventeenth century, was fundamental in shaping the subsequent development of Western society. Newton, who was seeking God in Nature as in the Scriptures, paradoxically left behind a world in which religion is no longer central. As I wrote in *Les Bâtisseurs du ciel* (Builders of the Heavens),

> *with [Newton], science helped break down the constraints imposed by faith which had hitherto dominated Western thought. Thus liberated, science has taken humanity on a journey to the edge of the universe, describing for us the touching fragility of our tiny planet.*[29]

The eighteenth century saw the triumph of Newton's theories: a perfectly mathematized celestial mechanics capable of predicting the return of Halley's Comet and the details of the lunar orbit. It was in this century that we discovered the first planet invisible to the naked eye (Uranus), the first catalog of nebulae was prepared, and we could conceive an invisible star, a precursor to the black holes predicted by general relativity. In terms of culture, it was also the century of salons and enlightened intellectuals at which one could hear Voltaire, Diderot, d'Alembert, and other philosophers hold forth with ease about science, society, morals, and politics. Unfortunately, the vast majority of our philosophers and writers in the present era are unable to competently discuss the big bang, superconductivity, or quantum gravity. The cause of this tragic disparity is the chasm that opened between the sciences and the humanities during the nineteenth century. One of the (probably unrealistic) goals of my work as a researcher and writer is in part to mend the lost links between science, philosophy, art, and literature.